\documentclass
[twocolumn,showpacs,preprintnumbers,amsmath,amssymb,prc]{revtex4}
\usepackage{graphicx}
\usepackage{dcolumn}
\usepackage{bm}

\begin{document}
\title{Estimates for temperature in projectile like fragment in geometric and
transport models}

\author{S. Mallik$^1$, S. Das Gupta$^2$ and G. Chaudhuri$^{1}$}

\affiliation{$^1$Theoretical Physics Division, Variable Energy Cyclotron Centre, 1/AF Bidhan Nagar, Kolkata 700064, India}
\affiliation{$^2$Physics Department, McGill University, Montr{\'e}al, Canada H3A 2T8}

\date{\today}

\begin{abstract}
Projectile like fragments emerging from heavy ion collision have an
excitation energy which is often labeled by a temperature.  This
temperature was recently calculated using a geometric model.  We expand
the geometric model to include also dynamic effects using a transport model.
The temperatures so deduced agree quite well with values of temperature
needed to fit experimental data.
\end{abstract}

\pacs{25.70Mn, 25.70Pq}

\maketitle
\section{Introduction}
Projectile multifragmentation is a practical tool for producing exotic nuclei
in the laboratory and remains a very active field of research both
experimentally and
theoretically.  This theoretical paper deals with one aspect of projectile
fragmentation.

There are many theoretical models for projectile multifragmentation:
we will not try to give an exhaustive list.  A few are:
the statistical multifragmentation model(SMM) \cite{Bondorf} (see also
\cite{Ogul} and \cite{Botvina} for application of SMM to projectile
fragmentation),;
heavy ion phase space exploration (HIPSE) model \cite{Lacroix} (see also
\cite{Mocko} for an application); antisymmetrised molecular dynamics (AMD)
model \cite{Ono} (see also \cite{Mocko} for applications); the abrasion-
ablation model of Gaimard-schmidt-Brohm \cite{Gaimard,Brohm}, the EPAX\cite
{Summerer} model and others.

In recent times we proposed a model \cite{Mallik2, Mallik3, Mallik101}
for projectile fragmentation whose predictions were compared with many
experimental data with good success. In contrast with the models mentioned above
our model uses the concept of temperature.  The concept of temperature is quite
familiar in heavy ion physics, whether to describe the physics of
participants (where
the temperature can be very high) or the physics of spectators (where the
temperature is expected to be much lower).  The ``nuclear caloric curve'' was
much researched as a signature of phase transition in nuclear systems \cite{
Pochodzalla,Dasgupta}.  Temperature of an emitting zone is often computed
using the ``Albergo'' formula \cite{Albergo}.
Thus temperature is a useful concept in projectile
spectator physics.

Our model has three parts.  To start
with we need an abrasion cross-section.  For a given impact parameter, this was
calculated using straight line trajectories for the projectile and the target
leading to a definite mass and shape to the projectile like fragment (PLF).
The PLF created will not be at zero temperature.
Let us label the mass of the PLF by $A_s(b)$, the mass of the projectile by
$A_0$.  It was conjectured that that the temperature of the PLF is a
universal function of the wound $1.0-A_s/A_0$.  This was parameterized as \cite
{Mallik3}:
\begin{equation}
T=7.5MeV-[A_s/A_0]4.5MeV
\end{equation}
A select set of experimental data from Sn on Sn collisions \cite{Ogul} were
used in
\cite{Mallik3} to arrive at the numbers above.  The formula was seen
to give very reasonable fits for many experimental data not only for Sn on Sn
 but other pairs of ions also.
The objective of the present work is solely to
investigate if we can arrive at the numbers generated by this simple
parametrization from a microscopic theory.  We may call this temperature
the primordial temperature.  The complete model does not stop here of course
and many more steps are needed to calculate observables.
We then postulate that this hot nuclear system will expand and break up into
all possible composites dictated solely by phase space.  This is the
canonical thermodynamic model (CTM) \cite{Das}.
The resulting hot composites will further evolve
by two-body sequential decays leading to the final products measured by
experiments.  Experimental results for many pairs of ions in the beam energy
range 140 MeV/n to 1 GeV/n were fitted quite well by this model
\cite{Mallik3,Mallik101}.

Our sole objective here is to estimate the value of the temperature of the
PLF when it is formed.  We estimated this in a geometric model \cite{SDG}.
As the numerical methods used for the geometric model will be extended to
a dynamical transport model, we need to review the geometric model first.
Conceptually the geometric model is simple but the calculations are
non-trivial.
We assume that the size and shape of the PLF is given by straight line cuts
and that divisions between participants and spectators are very clean
This excludes low beam energy.  Experiments at Michigan used 140MeV/n.
We made an ad-hoc assumption that we can use straight line cuts  at this
beam energy and higher; lower energy was not attempted.
In the geometric model, some parts of the projectile are removed,
which leaves the PLF with a crooked shape.  Nuclear structure effects
ascribe to this shape an excitation energy.  We now use the CTM \cite{Das}.
In that model for a given mass and temperature one can compute the excitation
energy per nucleon.  We reverse the procedure to go from excitation energy to
temperature.  Note that in the geometric model
the beam energy does not enter the calculation, the only assumption
being  that it is large enough for straight line trajectories to be valid.

In later sections we try to estimate the PLF temperature from a transport
model Boltzmann-Uehling-Uhlenbeck (BUU) calculation.  These are
the principal results of this work.  These calculations can
be used for many purposes but we restrict ourselves only to the objective
of trying to deduce a temperature for the PLF.

\section{Modeling the ground states of the colliding nuclei}
We start by choosing an impact paramater and boost one nucleus in its
ground state towards the other nucleus also in its ground state.  Throughout
this work, semi-classical physics is used.  We use Thomas-Fermi(TF)
solutions for ground states.  Complete details of our
procedure for TF solutions plus the choice of the interactions are given in
Ref. \cite{Lee}.   The kinetic energy density is given by
\begin{equation}
T(\vec r)=\int d^3p f(\vec{r},\vec{p})p^2/2m
\end{equation}
where $f(\vec{r},\vec{p})
$ is the phase space density.
Since we are looking for lowest energy we take, at
each $\vec{r}$, $f(\vec {r},\vec{p})$ to be non-zero from 0 to some maximum
$p_F(\vec{r})$.  Thus we will have
\begin{equation}
f(r,p)=\frac{4}{h^3}\theta [p_F(r,p)-p]
\end{equation}
The factor 4 is due to spin-isospin degeneracy and using the spherical symmetry of the TF solution we have dropped the vector sign on $r$ and $p$.
This leads to
\begin{equation}
T=\frac{3h^2}{10m}[\frac{3}{16\pi}]^{2/3}\int \rho(r)^{5/3}d^3r
\end{equation}
For potential energy we take
\begin{eqnarray}
V=A\int d^3r\frac{\rho^2(r)}{2}+\frac{1}{\sigma+1}B\int \rho^{\sigma+1}({r})d^3r\nonumber\\
+\frac{1}{2}\int d^3rd^3r' v(\vec{r},\vec{r}')\rho(\vec{r})\rho(\vec{r'})
\end{eqnarray}
The first two terms on the right hand side of the above equation are zero range
 Skyrme interactions.  The third which is a finite range term is often
suppressed and the
constants $A,B,\sigma$ are chosen to fit nuclear matter equilibrium density,
binding energy per nucleon and compressibility.  In heavy ion collisions, for
most
purposes, this will be adequate but for what we seek here, possibly a small
excitation energy, this is wholly inadequate.  Thomas-Fermi solution is obtained by minimizing $T+V$.  With only zero range force, $\rho(r)$ can be taken to be a constant which goes  abruptly to zero at some $r_0$ fixed by the total number of nucleons.  Now if $\rho$ is chosen to minimize the energy then, a nucleus, at this density with a cubic
shape is as good as a spherical nucleus.  Besides the minimum energy nucleus will have a sharp edge, not a realistic density distribution.  This problem does not arise in quantum mechanical treatment with Skyrme interaction.  Including a finite range potential in TF one recovers a more realistic density distribution for the ground state and one regains the nuclear structure effects which will contribute to excitation the PLF.  This is discussed in more detail
in Ref. \cite{Lee}.

Thomas-Fermi solutions for relevant nuclei were constructed with following force parameters.  The constants $A,B,$ and $\sigma$ (Eq.4) were taken to be
$A$=-1533.6 MeV fm$^3$, $B$=2805.3 MeV fm$^{7/2}$, $\sigma=7/6$.  For the finite range potential we chose an Yukawa :$V_y$.
\begin{equation}
V_y=V_0\frac{e^{-|\vec{r}-\vec{r'}|/a}}{|\vec{r}-\vec{r'}|/a}
\end{equation}
with $V_0$=-668.65 MeV and $a$=0.45979 fm.  Binding energies and density profiles for many finite nuclei with these parameters (and several others) are given in Ref. \cite{Lee}.  These have been used in the past
to construct TF solutions which collide in heavy ion collisions \cite{Gallego}.

\section{Methodology}
We use the method of test particles to evaluate excitation energies of a PLF with  any given shape.  The method of test particles is well-known from use of BUU
models for heavy ion collisions \cite{Bertsch}.  Earlier applications were made by Wong \cite{Wong}.  We will use the method of test particles for the
geometric model and as well as for BUU calculations in the later sections.

We first construct a TF solution using iterative techniques \cite{Lee}.  The TF phase space distribution will then be modeled by choosing test particles with appropriate positions and momenta using Monte-Carlo.  In most of this
work we consider 100 test particles ($N_{test}=100$) for each nucleon.  For example, the phase space distribution of $^{58}$Ni is described by 5800 test particles.  A PLF
can be constructed by removing a set of test particles.  Which test particles will be removed depends upon collision geometry envisaged.  For example, consider central collision of $^{58}$Ni on $^9$Be.
Let $z$ to be the beam direction.  For impact parameter $b$=0 we remove all test particles in $^{58}$Ni whose distance from the center of mass of $^{58}$Ni has  $x^2+y^2<{r_9}^2$ where $r_9=2.38$ fm is the half radius of $^9Be$.  The cases of non-zero impact parameter can be similarly considered.

We want to point out that this procedure of removing test particles from the
projectile may produce an error if the target is small and/or for very
peripheral collisions even if both the target and the projectile are heavy.
There can be transparency when small amounts of nuclear matter are traversed.
However this prescription of removing test particles from the projectile
when they are in the way of the target produces very definite predictions.
The transparency problem is treated well in the transport model that we will
get into later.

Continuing with the geometric model, we assume that
the shape and momentum distribution of the PLF can be described by removing some test particles as described above.  Of course this PLF will undergo many more changes later but all we are concerned with is the energy of the system at the time of "separation".  Since the PLF now is an isolated system, the energy will be conserved.  Of course the Coulomb force from the participants will continue to be felt by the PLF.  But the major effect of this will be on overall translation of the PLF and all we are interested in is intrinsic energy.

We now describe how we calculate the energy of this "crooked" shape object.  The mass number of the PLF is the sum of the number of test particles remaining divided by $N_{test}$.  Similarly the total kinetic energy of the PLF is the sum of kinetic energies of the remaining teat particles divided by $N_{test}$.  Evaluating
potential energy requires much more work.  We need a smooth density to be generated by positions of test particles.
We use the method of Lenk and Pandharipande to obtain this smooth density.  Other methods are possible \cite{Bertsch}.
Experience has shown that Vlasov propagation with Lenk-Pandharipande
prescription
conserves energy and momenta very well \cite{Lenk}.  For the geometric model
time propagation is not needed.  We will need that for BUU calculations in
later sections.

The configuration space is divided into cubic lattices. The lattice points are $l$ fm apart.  Thus the configuration space is discretized into boxes of size
$l^3$fm$^3$.  Density at lattice point $r_{\alpha}$ is defined by
\begin{equation}
\rho_L(\vec{r}_\alpha)=\sum_{i=1}^{AN_{test}}S(\vec{r}_{\alpha}-\vec{r}_i)
\end{equation}
The form factor is
\begin{equation}
S(\vec{r})=\frac{1}{N_{test}(nl)^6}g(x)g(y)g(z)
\end{equation}
where
\begin{equation}
g(q)=(nl-|q|)\Theta (nl-|q|)
\end{equation}
The advantage of this form factor is detailed in \cite{Lenk} so we will not enter that discussion here.  In this work we have always used $l$=1 fm and $n$=1

It remains to state how we evaluate the potential energy term (eq.(5)).  The zero range Skyrme interaction contributions are very simple.  For example the first term is calculated by using
\begin{equation}
A\int d^3r\frac{\rho^2(r)}{2}=A\sum_\alpha (l^3)\rho_L^2(r_{\alpha})/2
\end{equation}
With our choice $l^3$=1 fm$^3$.
For the third term in eq.(5) (the Yukawa term) is rewritten as $1/2\sum_{\alpha} \rho_L(\vec {r_{\alpha}}) \phi_L(\vec {r_{\alpha}})$ where $\phi(\vec{r})$ is the potential
at $\vec{r}$ due to the Yukawa, i.e.,$\phi(\vec{r})=\int V_y(|\vec{r}-\vec{r'}|)\rho(r')d^3r'$.

The calculation of Yukawa (and/or Coulomb) potential due to a charge distribution which is specified at points of cubic lattices is very non-trivial and involves iterative procedure.  This has been used a
great deal in applications involving time-dependent Hartree-Fock theory.  We will just give the references \cite{Koonin, Press, Varga}. We also found an unpublished MSU report very helpful \cite{Feldmeier}.

With this method we can calculate the total energy of the PLF.  However we are interested in excitation energy of the system which requires us to find
the ground state energy of the PLF which has lost some nucleons
from the projectile.  We can use TF theory to find this.  The iterative TF solution also gives the ground state energy.  But then we will be using two different methods
for evaluating energies, one for the PLF crooked shape and a different one for the PLF ground state.
Although the results are quite close, it is more consistent to use the same prescription for ground state energy and excitation energy. Hence, even for the ground state energy we generate test particles and go through the same procedure as for the PLF with crooked shape.

However we are not finished yet.  The canonical thermodynamic model (CTM) uses temperature.  However for a given temperature, mass number and charge, the average excitation energy per nucleon can be can be calculated \cite{Das}.  Given an excitation energy this can be used to deduce a temperature for the PLF
mass.  Results of the geometric model can be found in a recent publication
\cite{SDG}.  We will show some reasults in next two sections.

\section{Transport Model : Calculations I}
\begin{figure} [h]
\includegraphics[width=3.0in,height=3.0in,clip]{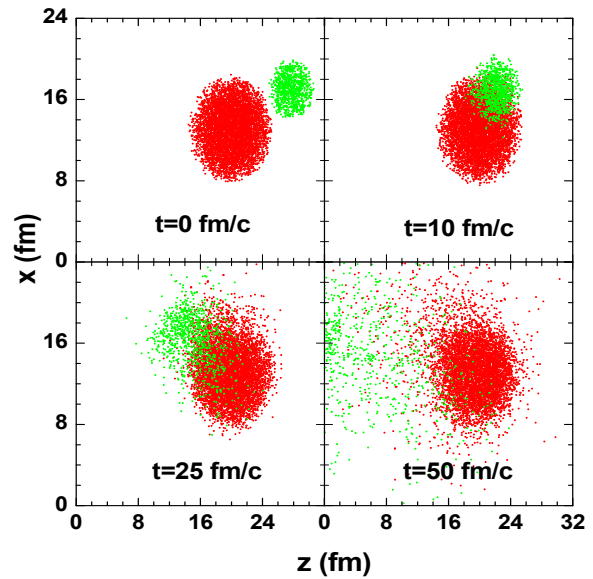}
\caption{Time evolution of $^{58}$Ni (red) and $^{9}$Be (green) test particles for $140$ MeV/nucleon at an impact parameter $b=4$ fm.}
\end{figure}
We begin transport model calculations to
identify and investigate properties of PLF.
Two nuclei, in their Thomas-Fermi ground states are boosted towards each
other with appropriate velocities at a given impact parameter.
We choose to do first $^{58}$Ni on $^9$Be
which was experimentally investigated at
MSU with beam energy 140MeV/n.  The calculations here follow the guidelines of
\cite{Bertsch} but some details were altered.  Two-body collisions are done
as in Appendix B of \cite{Bertsch} except that pion channels are closed as here
we are interested only in spectator physics (pions are created in the
participants) and besides the beam energy is low.  The mean field is
that prescribed in section II: zero range Skyrme plus the Yukawa of eq.(6).
The potential energy density is
\begin{equation}
v(\rho(\vec{r}))=\frac{A}{2}\rho^2(\vec{r})
+\frac{B}{\sigma+1}\rho^{\sigma+1}(\vec{r})+\frac{1}
{2}\rho(\vec{r})\phi(\vec{r})
\end{equation}
where $\phi(\vec{r})$ is the potential generated by the Yukawa:

$\phi(\vec{r})=\int V_y(|\vec{r}-\vec{r'}|\rho(\vec{r'})d^3r'$.
The Vlasov part is done as
in equations 2.14a and 2.14b of \cite{Lenk}.
\begin{equation}
\dot{\vec{r}}_i=\frac{\partial H}{\partial \vec{p}_i}=\frac{\vec{p}_i}
{m}
\end{equation}
\begin{equation}
\dot{\vec{p}}_i=-N_{test}\sum_{\alpha}\frac{\partial V}{\partial \rho_{\alpha}}
\vec{\nabla}_i\rho_{\alpha}
\end{equation}
where $V$ is the total potential energy of the system.
The Lenk-Pandharipande method
is a must here as all other known (to us) methods have numerical uncertainties
in energy evaluation which can hide the effects we are after.  The number of
test particles to represent the phase space is 100 per nucleon. Position and momenta of test particles are updated in time steps of 0.5 fm/c.
\begin{figure} [h]
\includegraphics[width=2.75in,height=2.75in,clip]{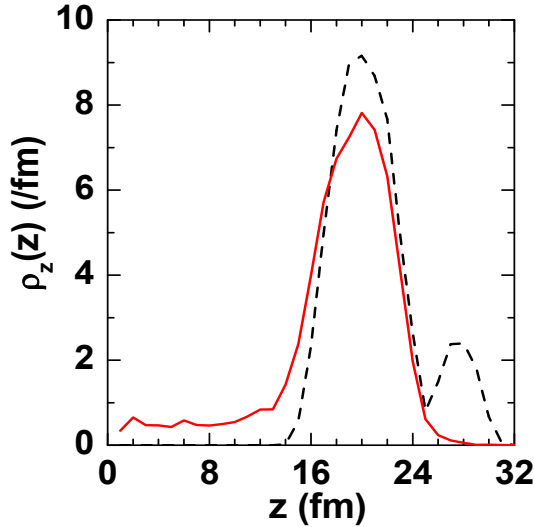}
\caption{$\rho_z(z)$ variation with z at t=0 fm/c (black dashed line) and 50 fm/c (red solid line) for $140$ MeV/nucleon $^{58}$Ni on $^{9}$Be reaction studied at an impact parameter $b=4$ fm.}
\end{figure}

We exemplify our method with collision at impact parameter b=4 fm.  It is
useful to work in the projectile frame and set the target nucleons with
the beam velocity in the negative z direction.  Fig.1 shows the test
particles at t=0 fm/c (when the nuclei are separate), t=10 fm/c, t=25 fm/c
and t=50 fm/c (Be has traversed the original Ni nucleus).  The calculation was
started with the center of Ni at 25 fm; at the end a large blob remains
centered at 25.  Clearly this is the PLF.  However a quantitative estimate
of the mass of the PLF and its energy requires further analysis.  This
type of analysis was done for each pair of ions and at each impact parameter
and details vary from case to case.  We exemplify this in one case only.

\begin{figure} [h]
\includegraphics[width=3.0in,height=2.75in,clip]{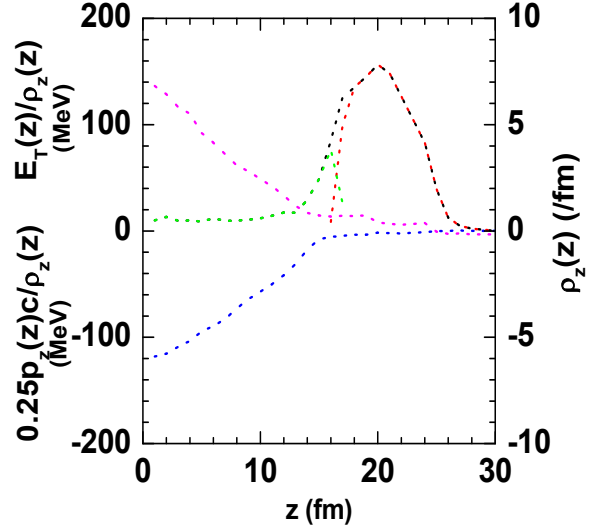}
\caption{Momentum per nucleon $p_zc(z)/\rho_z(z)$ (blue dashed line) total energy per nucleon $E_T(z)/\rho_z(z)$ (magenta dashed line) for $140$ MeV/nucleon $^{58}$Ni on $^{9}$Be reaction at an impact parameter $b=4$ fm studied at $t=50$ fm/c. Total density ($\rho^{t}_z(z)$), participant density ($\rho^{p}_z(z)$) and remaining part density ($\rho^{r}_z(z)$) along z-direction at $t=50$ fm/c are shown by black, red and green dashed lines respectively. For drawing all quantities in the same scale, $p_zc(z)/\rho_z(z)$ is divided by a factor 4.}
\end{figure}
For the analysis, it is convenient to introduce a kinetic energy density and
a z component of momentum density (we will use $p_zc$ rather than $p_z$).
Density at lattice point $r_{\alpha}$ is defined by
\begin{equation}
\rho_L(\vec{r}_\alpha)=\sum_{i=1}^{AN_{test}}S(\vec{r}_{\alpha}-\vec{r}_i)
\end{equation}
For kinetic energy density we use
\begin{equation}
T_L(\vec{r}_\alpha)=\sum_{i=1}^{AN_{test}}T_iS(\vec{r}_{\alpha}-\vec{r}_i)
\end{equation}
where $T_i$ is the kinetic energy of the i-th test particle.  It is also
useful to introduce a density for the the z-th component of momentum.
\begin{equation}
(p_zc)_L(\vec{r}_\alpha)=\sum_{i=1}^{AN_{test}}(p_zc)_iS(\vec{r}_{\alpha}-\vec{r}_i)
\end{equation}
The symbol $\alpha$ stands for values of the 3 co-ordinates of the lattice point
$\alpha=(x_l,y_m,z_n)$.
We will often, for a fixed value of $z_n$, sum over $x_l,y_m$.  For
example $\sum_{l,m}\sum_{i=1}^{AN_{test}}S((x_ly_mz_n)-\vec{r}_i)$ will be denoted
by $\rho_z(z_n)$.  Similarly for kinetic energy or total energy density:
$T(z_n)$ or $E_T(z_n)$.  Similarly for $p_zc(z_n)$ etc.
\begin{figure} [h]
\includegraphics[width=2.75in,height=2.75in,clip]{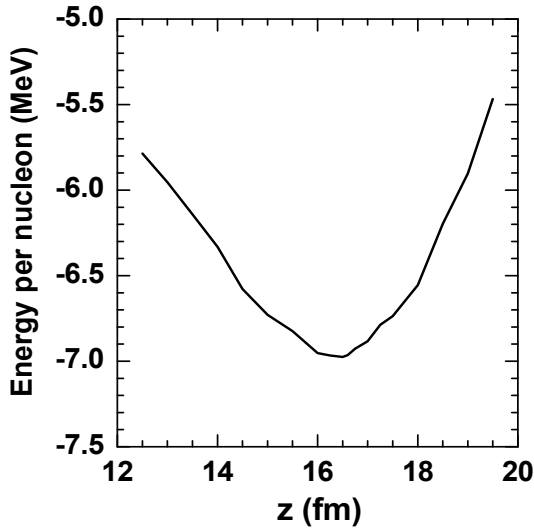}
\caption{Energy per nucleon of the test particles remains right side of the separation (z) for $140$ MeV/nucleon $^{58}$Ni on $^{9}$Be reaction at an impact parameter $b=4$ fm studied at $t=50$ fm/c.}
\end{figure}

 In Fig. 2 we plot $\rho_z(z)$ as a function of z at t=0 (when the the nuclei
start to approach each other) and at t=50 fm/c (when Be has traversed Ni).
Fig. 3 adds more details to the situation at 50 fm/c.  At far right one
has the PLF.  Progressively towards left one has the participant zone
characrised by a higher energy per nucleon $E_T(z)/\rho_z(z)$ and lower value
of $p_zc$ per nucleon (=$p_zc(z)/\rho_z(z)$).
Closer to the left edge one has target spectators.  In order to specify
the mass number and energy per nucleon of the PLF we need to specify
which test particles belong to the PLF and which to the rest (participant and
target spectators).  Our configuration box stretches from z=0 to z=33 fm.
If we include all test particles in this range we have the full system with
the total particle number 67(58+9) and the total energy of target plus projectile
in the projectile frame. Let us consider constructing a wall at z=0 and
pulling the wall to the right.  As we pull we leave out the test particles
to the left of the wall.  With the test particles to the right of the wall
we compute the number of nucleons and the total energy per nucleon.  The
number of particles goes down and initially the energy per nucleon will go
down also as we are leaving out the target spectators first and then the
participants.  At some point we enter the PLF and if we pull a bit further
we are cutting off part of the PLF giving it a non-optimum shape.  So
the energy per nucleon will rise.  The situation is shown in Fig.4.  The point
which produces this minimum is a reference point.  The test particles to
the right are taken to belong to PLF; those, to the left are taken to
represent the participants and target spectators. Not surprisingly, this point
is in the neighbourhood where both $E_T(z)/\rho(z)$ and
$p_zc/\rho(z)$ flatten out.

\begin{figure} [h]
\includegraphics[width=2.75in,height=2.75in,clip]{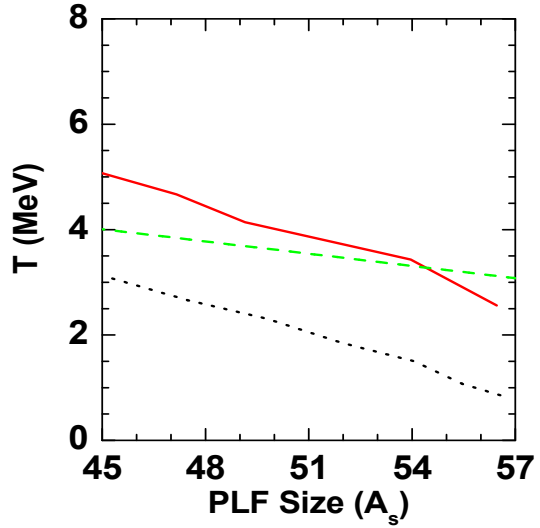}
\caption{Temperature profile obtained from BUU model calculation (red solid line) for $140$ MeV/nucleon $^{58}$Ni on $^{9}$Be reaction compared with that calculated from general formula (green dashed line) and geometrical model (black dotted line).}
\end{figure}

\begin{figure} [h]
\includegraphics[width=2.75in,height=2.75in,clip]{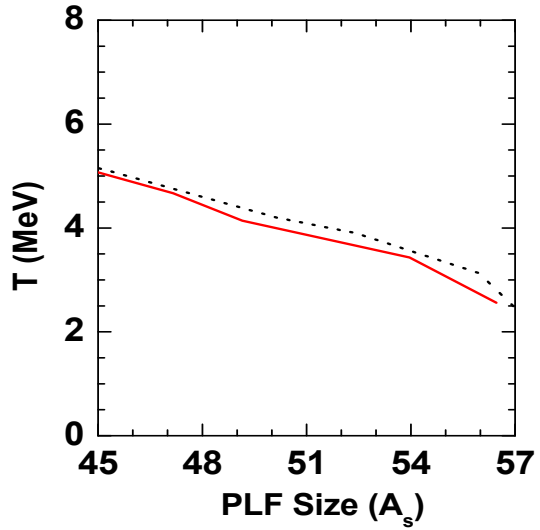}
\caption{Temperature profile obtained from BUU model calculation for $^{58}$Ni on $^{9}$Be reaction studied at $140$ MeV/nucleon (red solid line) and $400$ MeV/nucleon (black dotted line).}
\end{figure}

In Fig. 5 we to compare BUU results for $^{58}$Ni on $^9$Be at 140 MeV/n
with results from the general formula eq.(1) and the geometrical model.
As conjectured in \cite{SDG}, the geometrical model temperatures are driven up
when dynamics is included.  In Fig.6 we have compared results at 140 MeV/n
with results at 400 MeV/n.  We are not aware of any experiments at 400 MeV/n,
this was done merely to check if in BUU, PLF physics is sensitive to beam
energy.  Geometrical model assumes it is not.  Fig. 7 shows results from BUU
calculations for $^{40}$Ca on $^9$Be at 140 MeV/n.

\begin{figure} [h]
\includegraphics[width=2.75in,height=2.75in,clip]{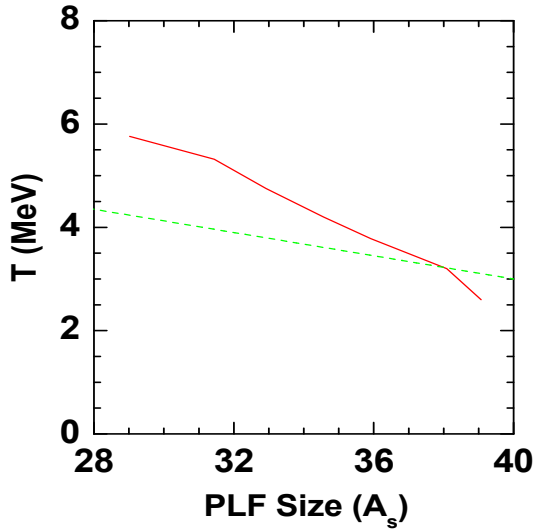}
\caption{Temperature profile obtained from BUU model calculation (red solid line) for $140$ MeV/nucleon $^{40}$Ca on $^{9}$Be reaction compared with general formula (green dashed line).}
\end{figure}

\section{Transport Model : Calculations II}
Vlasov propagation with Skyrme plus Yukawa for large ion collisions is not
practical.  Given nuclear densities  on lattice points, one is required to
to generate the potential which arises from the Yukawa interaction.  Standard
methods require iterative procedures involving matrices.  In the case of Ni
on Be, in the early times of the collision, the matrices are of the order of
1000 by 1000: as the system expands the matrices grow in size reaching about
7000 by 7000 at t=50 fm/c.  If we want to do large systems (Sn on Sn for
example) very large computing efforts are required.

To treat large but finite systems we use a mean field Hamiltonian
Lenk and Pandharipande devised for finite nuclei.  The mean field involves
not only the
local density but also the derivative of local density upto second
order.  The derivative terms do not affect nuclear matter properties but
in a finite system it produces quite realistic diffuse surfaces and
liqid-drop binding energies.

In order to keep the same notation as used in previous sections, we write
the Lenk-Pandharipande mean field as follows.The mean field potential is
\begin{equation}
u(\rho(\vec{r}))=A\rho(\vec{r})+B\rho^{\sigma}(\vec{r})+\frac{c}{\rho_0^{2/3}}
\nabla_r^2[\frac{\rho(\vec{r})}{\rho_0}]
\end{equation}
The potential energy density is
\begin{equation}
v(\rho(\vec{r}))=\frac{A}{2}\rho^2(\vec{r})+
\frac{B}{\sigma+1}\rho^{\sigma+1}(\vec{r}) +\frac{c\rho_0^{1/3}}{2}
\frac{\rho(\vec{r})}{\rho_0}\nabla_r^2[\frac{\rho(\vec{r})}{\rho_0}]
\end{equation}
Since there is no Yukawa term, the values of $A$ and $B$ and possibly
$\sigma$ need to be changed from the values used in section II (and the previous
section) to keep the property  of nuclear matter unchanged.  For calculations
reported in this  section the values are $A$=-2230.0 MeV $fm^3$,
B=2577.85 MeV $fm^{7/2}, \sigma=7/6$.  These values are taken from a previous
work \cite{Dasgupta}.  The value of the constant $\rho_0$ is
0.16 $fm^{-3}$ and the value of constant $c$ is -6.5 MeV.

Next problem is to find the ground state energy of a nucleus with $A$
nucleons.  Here we have used a variational method.  A parametrisation of
realistic density distribution was given by Myers which has been used many
times for heavy ion collisions \cite{Myers,Cecil,Lee}.   This parametrisation
is
\begin{equation}
\rho(r)=\rho_M[1-[1+\frac{R}{a}]\exp(-R/a)\frac{sinh(r/a)}{r/a}],  r<R
\end{equation}
\begin{equation}
\rho(r)=\rho_M[(R/a)cosh(R/a)-sinh(R/a)]\frac{e^{-r/a}}{r/a},  r>R
\end{equation}
There are two parameters here: ``$a$'' which controls the width of the surface
and $\rho_M $(or $R$) which controls the equivalent sharp radius.  The
distribution satisfies 4$\pi\int_0^{\infty}\rho(r)r^2dr=
A=\frac{4\pi}{3}\rho_MR^3$.  Thus no special normalization is required.  The
distribution has the advantage that equivalent sharp radius $R$ is simply
proportional to $A^{1/3}$ while the half-density radius of a Fermi distribution
does not have this simple proportionality.  Comparison with Fig.2 in
\cite{Lenk} shows that the energy calculated by this variational calculation
is quite close to what is given by Thomas-Fermi theory.
\begin{figure} [h]
\includegraphics[width=3.0in,height=3.0in,clip]{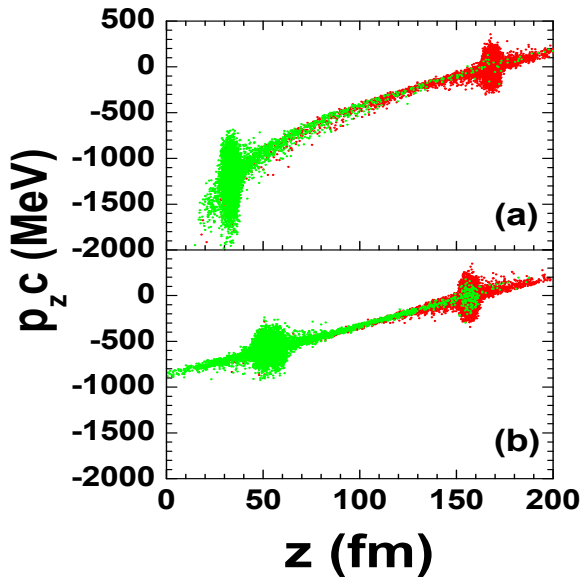}
\caption{$p_zc$ vs $z$ variation of projectile (red) and target (green) test particles at t=200 fm/c for $^{124}$Sn on $^{119}$Sn reaction studied at an impact parameter $b=4$ fm with energy (a) $600$ MeV/nucleon (relativistic kinematics) and (b)$200$ MeV/nucleon (non-relativistic kinematics).}
\end{figure}
\begin{figure} [h]
\includegraphics[width=2.75in,height=2.75in,clip]{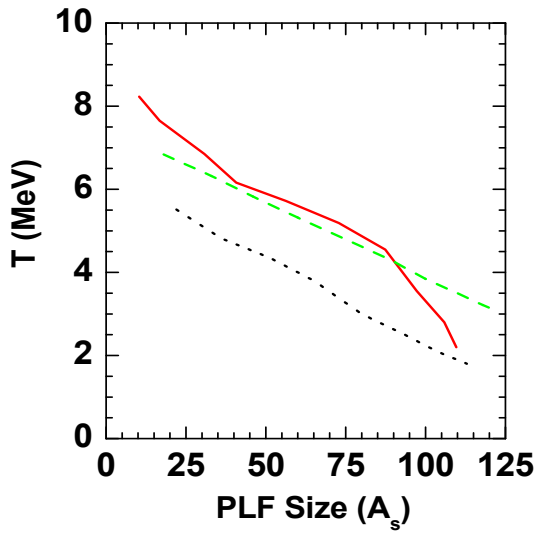}
\caption{Curve similar to Fig. 5 but for $^{124}$Sn on $^{119}$Sn reaction at 600 MeV/nucleon.}
\end{figure}

\begin{figure} [h]
\includegraphics[width=2.75in,height=2.75in,clip]{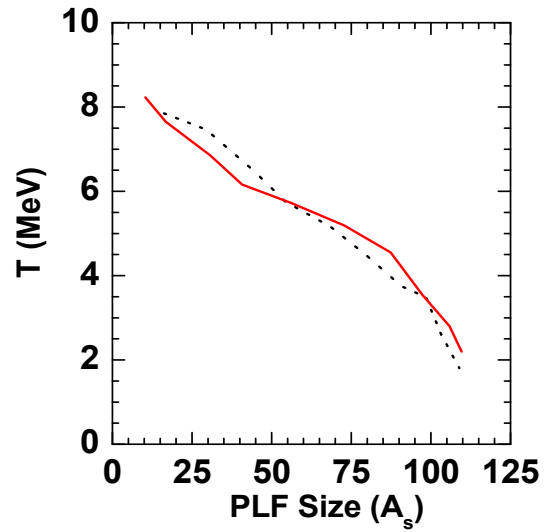}
\caption{Temperature profile obtained from BUU model calculation for $^{124}$Sn on $^{119}$Sn studied at $600$ MeV/nucleon using relativistic kinematics (red solid line) and $200$ MeV/nucleon using non-relativistic kinematics (black dotted line).}
\end{figure}

We do two cases of large colliding systems with the Lenk-Pandharipande
mean fields : $^{124}$Sn on $^{119}$Sn and $^{58}$Ni on $^{181}$Ta.
For these large colliding systems we reduced the number of test particles
per nucleon from 100 to 50; $N_{test}=50$.  Fig. 8 shows scatter of test
particles in the $z,p_zc$ plane
for Sn on Sn at time t=200 fm/c for beam energy (a) 600 MeV/n and (b) 200 MeV/n for
impact parameter 8 fm.  The plot, as before, is in the projectile frame
and identifies projectile like spectator, participant zone and target like
spectator.  In the 200 MeV/n calculations and all calculations in the previous
sections Vlasov propagation is non-relativistic but collisions are
treated relativistically (Appendix B of \cite{Bertsch}).  Experimental
data for $^{124}$Sn on $^{119}$Sn at 600 MeV/n are available \cite{Ogul}.
For 600 MeV/n beam energy, relativistic kinematics is used for propagation
of test particles.  This means the following.  In the rest frame of each nucleus
the Fermi momenta of test particles is calculated in the standard fashion
except that once they are generated we treat them like relativistic momenta.
Relativistic kinetic energy per nucleon in the rest frame of the nucleus,
on the average, becomes only slightly
different from  the non-relativistic value (about 0.3 MeV per nucleon).
As before, We work in the
rest frame of the projectile and the transformation of momenta of test
particles of the target to the projectile frame is relativistic.  In between
collisions, the test particles move with
$\dot{\vec{r}}=(\vec{p}c/e_{rel})c$ instead of $\vec{p}/m$.  Similarly the
change of momentum in test particles induced by the mean field is considered
to be the change in
relativistic momentum.  However these changes made little difference since
in the projectile frame the PLF test particles move slowly.

The 600 MeV/n results are shown in Fig.9.  Calculations for 200 MeV/n are
compared with the 600 MeV/n in Fig. 10.  Results are very similar.  This
fulfills the postulate of limiting fragmantation.  Fig. 11 gives results
for $^{58}$Ni on $^{181}$Ta where the beam energy is 140 MeV/n.  Experiment
at this energy was done at Michigan State University.

\begin{figure} [h]
\includegraphics[width=2.75in,height=2.75in,clip]{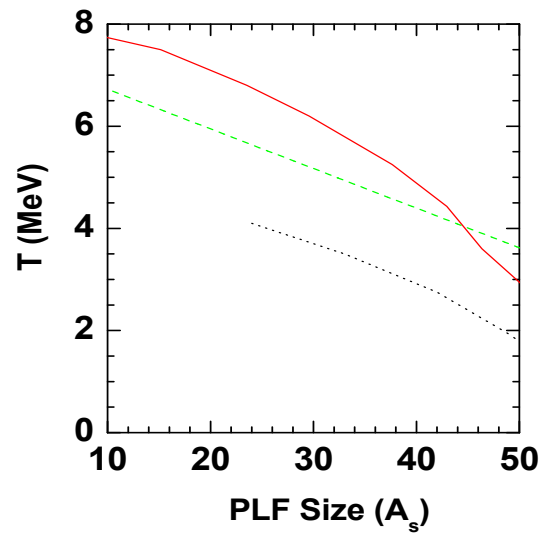}
\caption{Curve similar to Fig. 5 but the target is $^{181}$Ta instead of $^9$Be.}
\end{figure}
\section{Discussion}
It is quite gratifying that detailed BUU calculations bear out the two
striking features  of temperature profile in the PLF.  These are : (a)
temperatures are of the order of 6 MeV and (b) there is a very definitive
dependence on the intensive quantity $A_s/A_0$, temperature falling as this
increases.

For large ion collisions the PLF slows down slightly in the lab
frame (i.e., in the projectile frame it acquires a net small negative velocity).
The PLF is excited.  Comparison with the geometric model seems to confirm that
a large part of the excitation energy owes its origin to nuclear structure
effects.  The size of the PLF is also larger than what it would be if the PLF
were in ground state.  Although the shape was not analysed,
in general, a low density tail spreads out longer than in nuclei in their
ground state.
\section{acknowledgement}
Part of this work was done at the Variable Energy Cyclotron Centre in Kolkata.
S. Das Gupta thanks Dr. D. K. Srivastava and Dr. A. K. Chaudhuri for hospitality
during visit at Variable Energy Cyclotron Centre. S. Mallik likes to
acknowledge hospitality at McGill for a four months stay.  This work was
supported in part by Natural Sciences and Engineering Research Council of Canada.

\end{document}